\documentstyle[12pt,epsfig]{article}

\begin{document}
\begin{titlepage}
\title{Can higher order curvature theories explain rotation curves of galaxies ?}

\author{S. Capozziello\thanks{capozziello@sa.infn.it},
V.F. Cardone\thanks{winny@na.infn.it}, S.
Carloni\thanks{carloni@sa.infn.it},
A. Troisi\thanks{antro@sa.infn.it}\\
 {\em\small Dip. di Fisica "E.R. Caianiello", Universit\'a di Salerno} \\
 {\em \small and Istituto Nazionale di Fisica Nucleare, Sez. di Napoli, Gruppo Coll. di Salerno} \\
 {\em\small Via S. Allende, I-84081 Baronissi (Sa), Italy.}\\}

\date{\today}

\maketitle

\begin{abstract}

Higher order curvature gravity has recently received a lot of
attention due to the fact that it gives rise to cosmological
models which seem capable of solving dark energy and quintessence
issues without using ``ad hoc" scalar fields. In this letter, a
gravitational potential is obtained which differs from the
Newtonian one because of a repulsive correction increasing with
distance. We evaluate the rotation curve of our Galaxy and compare
it with the observed data in order both to test the viability of
these theories and to estimate the scalelength of the correction.
It is remarkable that the Milky Way rotation curve is well fitted
without the need of any dark matter halo and a similar result
tentatively holds also for other galaxies.

\end{abstract}

\thispagestyle{empty} \vspace{20.mm}
 PACS number(s): 04.50.+h, 98.80.-k, 98.35.df, 95.35.+d \\

\vspace{5.mm}

\vfill

\end{titlepage}

\maketitle

\section{Introduction}

In the last few years a very significant bulk of data have been
accumulated from different observational campaigns leading to an
unexpected picture of the universe. The Hubble diagram of type Ia
Supernovae (SNeIa) \cite{SNeIa} and the precise measurements of
the anisotropy spectrum of the cosmic microwave background
radiation (CMBR) by the BOOMERanG, MAXIMA and WMAP collaborations
\cite{CMBR} strongly suggest that the universe is spatially flat
and undergoing accelerating expansion driven by an unknown kind of
{\it dark energy}. In fact, when combined with the data for the
matter density parameter $\Omega_M$, these results lead to the
conclusion that the contribution $\Omega_X$ of the dark energy is
the dominant one, being $\Omega_M \simeq 0.3, \Omega_X \simeq
0.7$. On the other hand, astrophysical observations on galactic
scales indicate that the luminous matter is unable to give account
of the experimental rotation curves \cite{rcflat}. From a
theoretical point of view this phenomenon is usually referred to
the existence of the so called {\it dark matter}. Its nature is
still matter of debate with different explanations ranging from
baryonic MACHOs to cold dark matter particles \cite{dm}.

Recently, in the cosmological framework, much attention has been
devoted to higher order theories of gravity
\cite{capozcurv,noi1,noi2,CDTT,Vollick,Odintsov,Francaviglia}. The
reasons to take into account such teories come from fundamental
physics where higher order curvature terms in the effective action
of gravity are the results of interactions of scalar and graviton
fields on curved spacetimes \cite{Birrell}. Besides, in this
framework,  several successful inflationary models have been
formulated solving the shortcomings of cosmological standard model
\cite{Starobinsky,Suen}.

In \cite{capozcurv,noi1,noi2} and then in
\cite{CDTT,Vollick,Odintsov,Francaviglia}, it has been showed that
it is possible to obtain the observed accelerating dynamics of
universe expansion by taking into account higher order curvature
terms into the gravitational Lagrangian. Furthermore, in
\cite{noi2}, a successful test with SNeIa data has been performed.
Having tested such a scheme on cosmological scales, it is
straightforward to try to complement the approach by analyzing the
low energy limit of these theories in order to see whether this
approach is consistent with the {\it local} (i.e. on galactic
scale) physics. This is the aim of this letter where we study the
exterior space-time geometry generated by a spherically symmetric
point-like source, in the framework of higher order theories
assuming that the gravity Lagrangian contains a power law of Ricci
scalar (i.e. ${\cal{L}} \propto f_0 R^n$). We find that, in the
weak field limit, the Newtonian potential is modified by an
additive term which scales with the distance $r$ as a power law
depending on $n$. Having obtained the corrected gravitational
potential, we then theoretically evaluate the rotation curve of
our Galaxy and compare it with the observed data. This test shows
that the correction term allows to well fit the Milky Way rotation
curve without the need of dark matter. These results suggest that
considering higher order gravitational theories can provide both
an explanation to dark energy and dark matter.

\section{Exact vacuum solution near a spherically symmetric source}

We study the low energy limit of higher order theories to verify
if such a scheme is in agreement with astrophysical observations
on galactic scales. We refer to the field equations obtained in
\cite{capozcurv,noi1,noi2}. They are:

\begin{eqnarray} \label{3} G_{\alpha\beta} & = & \frac{1}{f'(R)} \Bigg \{ \frac{1}{2} g_{\alpha\beta} \left [
f(R) - R f'(R) \right ] + \nonumber \\
~ & ~ & + f'(R)^{;\mu \nu} \left ( g_{\alpha\mu} g_{\beta\nu} - g_{\alpha\beta} g_{\mu\nu} \right ) \Bigg \} \ ,
\end{eqnarray}
the prime denoting derivatives with respect to $R$. These field
equations can be decomposed into the time\,-\,time component and
the trace which can be combined in a simpler relation. Assuming
$f(R) = f_0 R^n$, as in \cite{noi1,capozcurv,noi2}, we obtain:

\begin{equation}
\label{14} R_{tt} = \left(\frac{2 n - 1}{6n}\right) \ g_{tt} R
\end{equation}
depending explicitly on the exponent $n$ of the theory.

Note that the off\,-\,diagonal Einstein equations reduces to
identities for the metric that we are  going to use here, which is

\begin{equation}
\label{15} ds^2 = - \frac{1}{K(r)} dt^2 + K(r) dr^2 + r^2 d\theta^2 +
 r^2 \sin{\theta} d\phi^2 \ ,
\end{equation}
where the spherical symmetry is assumed.
 Inserting it into
Eq.(\ref{14}), we get an equation for $K(r)$ that can be
analytically solved. We find\,:

\begin{equation}
K(r) = C_1 r^{\alpha(n)} + C_2 r^{\beta(n)} \ ,
\label{eq: kappa}
\end{equation}

\begin{equation}
\alpha(n) = \sqrt{\frac{4 n - 1}{2 (n - 1)}} \ \times \ \left [ {\cal{P}}(n) - {\cal{Q}}(n) \right ] \ ,
\label{eq: alpha}
\end{equation}

\begin{equation}
\beta(n) = \sqrt{\frac{4 n - 1}{2 (n - 1)}} \ \times \ \left [ {\cal{P}}(n) + {\cal{Q}}(n) \right ] \ ,
\label{eq: beta}
\end{equation}

\begin{equation}
{\cal{P}}(n) = - \frac{3}{2} \ \left ( \frac{3 n - 1}{4 n - 1} \right ) \ \sqrt{\frac{2 (4 n - 1)}{n - 1}} \ ,
\label{eq: pn}
\end{equation}

\begin{equation}
{\cal{Q}}(n) = \sqrt{{\cal{P}}^2(n) - 4} \ .
\label{eq: qn}
\end{equation}
We have now all what we need to obtain the gravitational potential
$\Psi(r)$ generated by a point-like source. To this aim, one only
needs to remember that it is $g_{rr} = 1 + 2 \Psi/c^2$, so that we
immediately get\,:

\begin{equation}
\Psi(r) = \frac{c^2}{2} \left [ C_1 r^{\alpha(n)} + C_2 r^{\beta(n)} \right ] \ .
\label{eq: psipoint}
\end{equation}
where we have opportunely chosen the additive arbitrary constant.
Up to now, the two integration constants $C_1$ and $C_2$ are
completely arbitrary, but some general considerations help in
choosing their values. First, we note that, in order to get the
correct physical dimensions for $\Psi$, $C_1$ and $C_2$ must have
the dimensions of $l^{-\alpha(n)}$ and $l^{-\beta(n)}$, being $l$
a length. It is, thus, convenient to redefine them as\,:

\begin{equation}
C_1 = \frac{s_1}{\xi_1^{\alpha(n)}} \ \ , \ \ C_2 = \frac{s_2}{\xi_2^{\beta(n)}}
\label{eq: conetwo}
\end{equation}
being $s_i$ the sign of $C_i$ and $(\xi_1, \xi_2)$ two
undetermined scalelengths. Let us now consider the behaviour of
the two powers $(\alpha, \beta)$ in Eq.(\ref{eq: psipoint}) with
respect to the exponent $n$. A careful analysis shows that\,:

\begin{displaymath}
{\rm for} \ \ n \in (0.25, 1) \ \ \alpha(n) \simeq -1 \ \ , \ \
\beta(n)
> 0 \ \ ;
\end{displaymath}

\begin{displaymath}
{\rm for} \ \ n \notin (0.25, 1) \ \ \alpha(n) < 0 \ \ , \ \
\beta(n) \simeq -1 \ \ .
\end{displaymath}
Let us take into account the case $n \in (0.25, 1)$. The first
term in Eq.(\ref{eq: psipoint}) is the leading one for small $r$,
i.e. for $r << \xi_2$. Actually, we know that on small scale (at
least for distances within the Solar System) the gravitational
potential is Newtonian, i.e. it is attractive and scales as $1/r$.
In order to recover this behaviour, we thus need that $s_1 = -1$,
while we have no constraints on $s_2$. The case $s_2 < 0$ may be
excluded, since the corresponding gravitational force between two
point masses should be more and more attractive as the bodies are
far away and this is not a physical case. We may thus conclude
that, for $n \in (0.25, 1)$, it is $s_1 = -1$ and $s_2 = +1$.

Similar considerations show that the case $n \notin (0.25, 1)$ is
unphysical. For these values of $n$, both terms in Eq.(\ref{eq:
psipoint}) are present on small scales. In order to recover the
Newtonian potential on such scales, we have to impose $s_2 = -1$.
On the other side, if $s_1$ were equal to 1, then we will obtain a
repulsive correction to the gravitational force on small scales
leading to a result contrasting with observations. The same
contradiction will be present if $s_1 = -1$ since, in this case,
the gravitational force should be more attractive than as probed
by many experiments. We have thus to conclude that models with $n
\notin (0.25, 1)$ are not physically acceptable since their low
energy limit does not reproduce the well known Newtonian potential
on Solar System scales.

From now on, we will only consider models with $n$ in the range
$(0.25, 1)$ and rewrite the gravitational potential of a pointlike
source with mass $M$ as\,:

\begin{equation}
\Psi(r) = - \frac{c^2}{2} \left [ \left ( \frac{r}{\xi_1} \right )^{-1} -
\left ( \frac{r}{\xi_2} \right )^{\beta(n)} \right ] \ .
\label{eq: psiend}
\end{equation}
A first estimate of $\xi_1$ may be obtained observing that, for $r << \xi_2$, Eq.(\ref{eq: psiend}) reduces to

\begin{displaymath}
\Psi(r) \sim - \frac{c^2}{2} \left ( \frac{r}{\xi_1} \right )^{-1} \ .
\end{displaymath}
Since we have to recover the Newtonian potential at these scales, we have to fix\,:

\begin{displaymath}
\xi_1 = \frac{2 G M}{c^2} \simeq 9.6 \times \frac{M}{M_{\odot}}
\times 10^{-17} \ {\rm kpc}\ ,
\end{displaymath}
with $M_{\odot}$ the mass of the Sun. The value of $\xi_2$ is a
free parameter of the theory. Up to now, we can only say that
$\xi_2$ should be much larger than the Solar System scale in order
not to violate the constraints coming from local gravity
experiments.

\section{The Milky Way rotation curve}

Eq.(\ref{eq: psiend}) gives the gravitational potential of a
pointlike source. Since real galaxies are not pointlike, we have
to generalize Eq.(\ref{eq: psiend}) to an extended source. To this
aim, we may suppose to divide the Milky Way in infinitesimal mass
elements, to evaluate the contribution to the potential of each
mass element and then to sum up these terms to get the final
potential. Assuming spheroidal symmetry and using Eq.(\ref{eq:
psipoint}), the gravitational potential generated by a galaxy
is\footnote{In this section, $R$ is the radial coordinate, not the
Ricci scalar.}\,:

\begin{equation}
\Psi(R, z) = \frac{c^2}{2} \left [ \Psi_1(R, z) + \Psi_2(R, z) \right ]\ ,
\label{eq: psitot}
\end{equation}
with

\begin{equation}
\Psi_i = C_i \int_{0}^{\infty}{R' dR' \int_{0}^{2 \pi}{ d\phi' \int_{-\infty}^{+\infty}{dz' \rho(R', z') r^{\eta_i}}}}\ ,
\label{eq: psiterms}
\end{equation}
being $\eta_i = -1 \ (\beta)$, for $i = 1 \ (2)$, while $\rho(R,
z)$ is the mass density of the galaxy (having assumed spheroidal
symmetry). We remark that we have adopted cylindrical coordinates
$(R, \phi, z)$ so that it is\,:

\begin{displaymath}
r^2 = | {\bf r} - {\bf r'} |^2 = R^2 + R'^2 - 2 R R' \cos{\phi'} + (z - z')^2 \ .
\end{displaymath}
In order to restore the correct dimensions of the potential, we
have to recast the two constants $C_1$ and $C_2$ in a way similar
to Eq.(\ref{eq: conetwo}). It is\,:
\begin{equation}
C_1 = - \frac{2 \ G}{c^2 \ \xi_1^{\alpha(n) + 1}} \ \ , \ \
C_2 = \frac{2 \ G}{c^2 \ \xi_2^{\beta(n) + 1}} \ \ .
\label{eq: conetwobis}
\end{equation}
A simple check shows that, with this assumption, the potential
$\Psi_i$ in Eq.(\ref{eq: psiterms}) has the correct physical
dimensions. Note also that, since $\alpha = -1$, the exact value
of $\xi_1$ is meaningless and hence $(n, \xi_2)$ are the only
parameters to determine the shape of the gravitational potential
for an extended source. The rotation curve may then be obtained by
the standard rule\,:

\begin{equation}
v_c^2(R) = R \left . \frac{\partial \Psi}{\partial R} \right |_{z = 0} \ .
\end{equation}
To test whether the theory is in agreement with observations and
to determine the parameter $\xi_2$, we have computed the Milky Way
rotation curve modeling our Galaxy as a two components system, a
spheroidal bulge and a thin disk. In particular, following
\cite{DB98}, we assume\,:

\begin{equation}
\rho_{bulge} = \rho_0 \left ( \frac{m}{r_0} \right )^{-1.8} \
\exp{\left ( - \frac{m^2}{r_t^2} \right )} \ , \label{eq:
rhobulge}
\end{equation}

\begin{equation}
\rho_{disk} = \frac{\Sigma_0}{2 z_d} \exp{\left ( - \frac{R}{R_d}
- \left | \frac{z}{z_d} \right | \right )} \label{eq: rhodisk}
\end{equation}
with $m^2 = R^2 + z^2/q^2$. The central densities $\rho_0$ and
$\Sigma_0$ are conveniently related to the bulge total mass
$M_{bulge}$ and the local surface density $\Sigma_{\odot}$ by the
following two relations\,:

\begin{displaymath}
\rho_0 = \frac{M_{bulge}}{4 \pi q \times 1.60851} \ ,\
\Sigma_0 = \Sigma_{\odot} \exp{\left ( \frac{R_0}{R_d} \right )}\ ,
\end{displaymath}
being $R_0 = 8.5 \ {\rm kpc}$ the distance of the Sun to the
Galactic Centre. Following \cite{DB98}, we fix the Galactic
parameters as follows\,:

\begin{displaymath}
M_{bulge} = 1.3 \times 10^{10} \ {\rm M_{\odot}} \ , \ r_0 = 1.0 \ {\rm kpc} \ , \ r_t = 1.9 \ {\rm kpc} \ ,
\end{displaymath}
\begin{displaymath}
\Sigma_{\odot} = 48 \ {\rm M_{\odot} \ pc^{-2}} \ , R_d = 0.3 R_0 \ , \ z_d = 0.18 \ {\rm kpc} \ .
\end{displaymath}
The Milky Way rotation curve $v_c(R)$ can be reconstructed
starting from the data on the observed radial velocities $v_r$ of
test particles. We follow \cite{DB98} reconstructing the rotation
curve from the data on H\,II regions and molecular clouds in
\cite{BB93} and those on a sample of classical Cepheids in the
outer disc obtained by Pont et al. \cite{Pont97}.

\begin{figure}
\centering \resizebox{8.5cm}{!}{\includegraphics{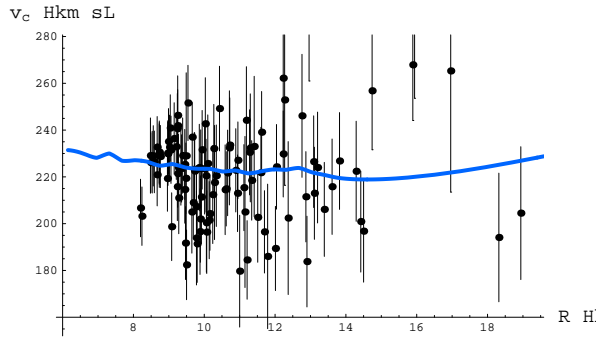}}
\caption{Observed data and theoretical Milky Way rotation curve
computed using the modified gravitational potential with $n =
0.35$ and $\xi_2 = 14.88$\,kpc. Note that the points with $R$
between 15.5 and 17.5 kpc are likely affected by systematic errors
as discussed in \cite{Pont97}.} \label{fig: rot}
\end{figure}

For a given $n$, we perform a $\chi^2$ test to see whether the
modified gravitational potential is able to fit the observed
rotation curve and to constrain the value of $\xi_2$. Since {\it a
priori} we do not know what is the range for $\xi_2$, we get a
first estimate of $\xi_2$ by a simple approach. For a given $R$,
we compute $\xi_2$ imposing that the theoretical rotation curve is
equal to the observed one. Then, we study the distribution of the
$\xi_2$ values thus obtained and evaluate both the median
$\xi_2^{med}$ and the median deviation $\delta \xi_2$. The usual
$\chi^2$ test is then performed with the prior that $\xi_2$ lies
in the range $(\xi_2^{med} - 5 \ \delta \xi_2, \xi_2^{med} + 5 \
\delta \xi_2)$. As a first test, we arbitrarily fix $n = 0.35$. We
get\,:

\begin{equation}
\xi_2 = 14.88 \ {\rm kpc} \ , \ \chi^2 = 0.96 \ \ .
\label{eq: bestxi}
\end{equation}
In Fig.\,\ref{fig: rot}, we show both the theoretical rotation
curve for $(n, \xi_2) = (0.35, 14.88)$ and the observed data. The
agreement is quite good even if we have not added any dark matter
component to the Milky Way model. This result seems to suggest
that our modified theory of gravitation is able to fit galaxy
rotation curves without the need of dark matter. As a final
remark, we note that $\xi_2^{med} = 14.37$\,kpc that is quite
similar to the best fit value. Actually, a quite good estimate is
also obtained considering the value of $\xi_2$ evaluated using the
observed rotational velocity at $R_0$. This suggest that a quick
estimate of $\xi_2$ for other values of $n$ may be directly
obtained imposing $v_{c,theor}(R_0; n, \xi_2) = v_{c,obs}(R_0)$.

\section{Conclusions}

In this letter, it has been analyzed the low energy limit of
higher order theories of gravity considering a power law function
of Ricci scalar for the gravitational Lagrangian. An exact
solution of the field equation has been obtained. The resulting
gravitational potential for a point-like source is the sum of a
Newtonian term and a contribute whose rate depends on a function
of the exponent $n$ of Ricci scalar. The potential agrees with
experimental data if $n$ ranges into the interval $(0.25, 1)$, so
that the correction term scales as $r^{\beta}$ with $\beta > 0$.

The following step is the generalization of this result to an
extended source as a galaxy. To this aim the experimental data and
the theoretical prediction for the rotation curve of Milky Way
have been compared. The final result has been that the modified
potential is able to provide a rotation curve which fits data
without adding any dark matter component. This result has to be
tested further before drawing a definitive conclusion against the
need for galactic dark matter. To this aim, one must show that a
potential like that predicted by our model is able to fit rotation
curves of a homogeneous sample of external galaxies with both well
measured rotation curves and detailed surface photometry. In
particular, the exponent $n$ coming out from the fit must be the
same for all the galaxies, while $\xi_2$ could be different being
related to the scale where deviations from the Newtonian potential
sets in.  On the other hand, however, it is worth  noting that for
$n = 1/3$, it is $\beta = 1$. Such kind of modified potential has
been taken into account (on the basis of a different theory) by
Mannheim \cite{Mannheim3} who has successfully fitted the rotation
curves of external galaxies. He has also found that the
normalization parameter (which plays in his theory the same role
as $\xi_2$) changes from one galaxy to another in agreement with
our expectations. This is an encouraging result since it states
that our model could be able to work correctly at least for one
value of the exponent $n$.

It is worth noting that modifications of the gravitational
potential are not the only way to fit galactic rotation curves
without adding any dark matter components. A quite successful
scheme is that adopted in the MOND theory \cite{MOND} where it is
the acceleration law to be changed. However, it is worth to stress
that our model is completely different from the MOND approach.
MOND has been originally proposed as a phenomelogical scheme to
solve an astrophysical problem and it cannot be derived from a
Lagrangian field theory (even if some tentatives have been done).
On the contrary, our modified gravitational potential is the
natural outcome of a Lagrangian theory and it is thus physically
well founded. Besides, contrary to MOND, our model has also a
cosmological counterpart \cite{noi1,capozcurv,noi2}.

In this paper, we have presented an indication that it is possible
to reduce the dark energy and dark matter issues under the same
higher order theories of gravity. In fact, this can give rise to
realistic models working at very large scales (cosmology) and
local scales (galaxies).

\end{document}